# Improving the Process-Variation Tolerance of Digital Circuits Using Gate Sizing and Statistical Techniques *

Osama Neiroukh[1] and Xiaoyu Song[2]
[1] M/S JF4-409, Intel Corporation, 2111 N.E. 25th Avenue, Hillsboro, OR 97124, USA
[2] Department of ECE, Portland State University, P. O. Box 751, Portland, OR 97207-0751, USA

## Abstract

*A new approach for enhancing the process-variation tolerance of digital circuits is described. We extend recent advances in statistical timing analysis into an optimization framework. Our objective is to reduce the performance variance of a technology-mapped circuit where delays across elements are represented by random variables which capture the manufacturing variations. We introduce the notion of statistical critical paths, which account for both means and variances of performance variation. An optimization engine is used to size gates with a goal of reducing the timing variance along the statistical critical paths. We apply a pair of nested statistical analysis methods deploying a slower more accurate approach for tracking statistical critical paths and a fast engine for evaluation of gate size assignments. We derive a new approximation for the max operation on random variables which is deployed for the faster inner engine. Circuit optimization is carried out using a gain-based algorithm that terminates when constraints are satisfied or no further improvements can be made. We show optimization results that demonstrate an average of 72% reduction in performance variation at the expense of average 20% increase in design area.*

## 1. Introduction

Recent advances in VLSI have continued to shrink device geometries at a steady rate in accordance with Moore's Law. However, this advancement has also been accompanied by increasing variations in the performance of fabricated circuits. Numerous factors have contributed to this trend including clock PLL jitter, noise, PV model inaccuracies, and manufacturing variations. Nevertheless, it is often desirable to manufacture ASICs on advanced technology nodes due to substantial increase in available device count, reduction in power consumption, higher yields and lower costs due to the larger 300mm wafers.

Researchers have recently focused on statistical analysis approaches in an attempt to grapple with these sources of performance variations. Statistical timing analysis models delay arcs as random variables and propagate timing constraints using probability distribution functions (pdfs). While a substantial focus has gone into the analysis aspect of this problem[1,2], recent research into statistical optimization of circuits has been surprisingly diminutive. Circuit optimization was done in [3] by using LANCELOT [4] but had severe limitation on circuit size and used unrealistic delay models. A concept of criticality of gates was used in [5] but did not address the variance of the timing path delays. A transistor level approach was presented in [6]. Several yield-specific techniques were presented in [7].

In this paper we present a unique approach that identifies worst negative statistical slack (WNSS) paths analogous to traditional worst negative slack (WNS) paths. Our method also provides flexibility for optimization objective function by assigning weights that enable user-driven tradeoffs between mean and variance of circuit performance.

The remainder of this paper is organized as follows:
- We present background on proposed research
- We formulate the problem of performance variability reduction in presence of statistical delays
- We derive a method for tracing the worst negative statistical slack (WNSS) path in a circuit
- We derive and demonstrate efficacy of a new approximation for quick calculation of the mean and variance of the maximum of random variables
- We present a robust gain-based sizing approach that handles a weighted sum of means and variances of delays
- Experimental results are presented and analyzed

## 2. Related work

### 2.1 Gate sizing

Gate sizing has been studied extensively in the literature. It is typically performed after technology mapping during logic synthesis and repeated throughout the design process. The aim of gate sizing is to assign sizes to gates in a circuit such that a performance objective function is satisfied.

* This research was supported by Intel Corporation.

Although sizing approaches relying on convex assumptions or analytical delay models have been proposed, more recent approaches tend to tackle the problem using greedy heuristics. According to [8], accurate delay models make gate sizing a non-linear, non-convex, constrained, discrete optimization problem. Most greedy gate sizing algorithms share several common elements [8, 9, 10, 11]. The critical path, sometimes referred to as the Worst Negative Slack (WNS) path, is usually targeted for optimization. We note that the WNS path can change as the optimization proceeds so the path being evaluated for resizing must be updated regularly during sizing iterations. The algorithms can be run in a constrained mode where delay for example is optimized first then area is recovered as far as possible without violating a delay constraint.

## 2.2 Statistical static timing analysis

The focus on use of statistical approaches in timing analysis is relatively new. Pioneering works in this field appeared in [12, 13, 14]. However, in the past few years statistical techniques for timing analysis of circuits have received tremendous focus with representative works including [15, 16, 17]. Static timing analysis relies on two operations for propagating timing through a network, sum and max. Performing these calculations on pdfs is more expensive computationally than their counterparts in the deterministic case. Moreover, the correlation between two pdfs needs be taken into account for accurate calculations.

## 3. Problem formulation and motivation

The starting point for our problem is a technology mapped digital circuit. Without loss of generality, this paper focuses on combinational circuits. We ignore interconnect delay though accounting for them can be readily accommodated. In fact, we postulate that our algorithm can help overcome the inherent interconnect uncertainty during pre-layout convergence by treating interconnect delays as random variables.

Our method uses discrete probability distribution functions (pdfs) throughout. A discrete pdf for random variable $X$ is defined as one or more points where $f(x) = P(X = x)$. The mean and variance of a discrete random variable are given by

$$\mu_X = \sum x_i f(x_i)$$
$$\sigma_X^2 = \sum (x_i - \mu_X)^2 f(x_i)$$

We assume that every gate delay in the circuit is represented by a normally distributed random variable which is consistent with the literature. Arrival times are propagated throughout the circuit as pdfs. We define the unconstrained timing variance minimization problem for a circuit as

$$\text{Minimize } \sigma_O^2$$

where

$\mu_O$ = Mean $(RV_O)$
$\sigma_O^2$ = Variance $(RV_O)$
$RV_O = \underset{i \in OUT}{Max} (RV_i)$ where the Max is the statistical Max operator on random variables
$RV_i$ = Random variable representing propagated arrival time of output $o_i$
$OUT = \{o_1, o_2, ..., o_N\}$ are the circuit's outputs

We note that the random variable $RV_O$ characterizes the mean and variance of the entire circuit. It should be highlighted that a circuit may have multiple outputs with close mean delays but different variances. In this case, all such outputs will contribute to the overall variance $\sigma_O^2$ of the circuit's performance. Alternatively, an output with the highest variance may have a much smaller mean than other outputs and reducing its variance will have minimal effect on overall variance of the circuit's performance. Any algorithm that attempts to alter $RV_O$ must account for both means and variances of delays simultaneously.

Fig. 1 gives a plot of $RV_O$ at different optimization points. The original line represents a pdf obtained by optimizing a circuit with a goal of minimizing the mean of the longest delay in the circuit. Such a circuit will typically exhibit the widest spread in performance due to high usage of smaller devices which exhibit more manufacturing variability. Depending on target application of circuit, such a performance variance around the center can represent undesirable uncertainty that should be minimized. In [18], reduction of uncertainty was shown to be a key strategy for designing leading edge industrial designs. Decreasing variance can increase the overall yield of a design. An example of this is optimization 1 in Fig. 1 which yields more functional units at period T relative to the original design. However, our technique is quite general and is not limited to yield maximization. Decreasing performance variance is also desirable on several other accounts even if it means relaxing the original timing targets. For example, circuits on the original curve to the left of "X" in Fig. 1 below will exhibit undesirable variance in power consumption due to both dynamic and leakage power variations. These variations in turn contribute uncertainties in thermal dissipation and reliability verification. The effects of such performance variations can adversely product qualification and time-to-market. In such instances, the 2[nd] optimization point shown below becomes desirable due to better tolerance to manufacturing variations. Our research is aimed at providing designers with a statistically aware gate sizing methodology that allows arbitrary tradeoffs between mean and variance of $RV_O$.

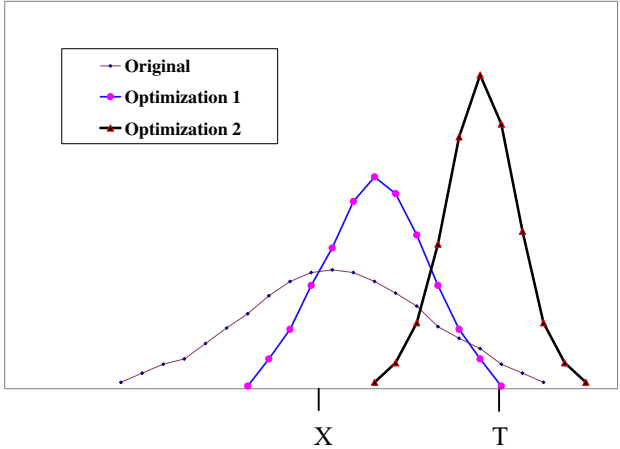

**Figure 1. Circuit Output Delay PDF**

## 4. Proposed approach

We studied several deterministic sizing techniques to evaluate their fitness as a basis for statistical sizing. Our preference for accurate gate delay models steered us away from methods [19,20,21], which require convex analytical expressions for gate delays. Such models not adequately capture the nonlinearities in current and foreseeable DSM technologies where manufacturing variations are prevalent. Our proposed approach is shown in Fig. 2. It builds on the deterministic algorithms presented in [8,11]. We show next how we deal with new challenges that arise when timing constraints are represented by random variables.

### 4.1 Pseudo code

```
Algorithm StatisticalGreedy
  repeat {
    FULLSSTA
    Trace critical (WNSS) path
    foreach g ∈ (gates on WNSS){
        extract subcircuit S around g
        SBestCost = Cost(S)
        GCurrentSize = CurrentSize(g)
        GBestSize    = GCurrentSize
        foreach I ∈ (sizes of g) {
            g in S ← I
            SNewcost = Cost(S)
            If(Snewcost < SBestCost) {
                GBestSize=I
                SBestCost=SNewCost
            }
        }
        If(GBestSize <> GCurrentSize)
            g.nextSize ← GBestSize
    }
    Resize scheduled gates
  } Until constraints met or no further improvement
Procedure Cost(Subcircuit S)
  Perform FASSTA on S
  Return ObjectiveFunction(S)
```

**Figure 2. Overview of StatisticalGreedy**

### 4.2 FULLSSTA

Our full statistical analysis engine is based on [15]. This approach discretizes pdfs at a user controlled sampling rate. We used 10-15 samples per pdf as a reasonable tradeoff between accuracy and speed. The operations sum and max are performed on discrete pdfs using shifting, scaling, and min/max reduction. In addition to propagating pdfs, we also calculate the mean and variance at every node and store these values for use in the fast timing engine (FASSTA). This component in our algorithm can be updated as needed to track the latest emerging research in statistical timing analysis and represents the outer loop for our iterations.

### 4.3 FASSTA

Statistical analysis methods such as FULLSSTA are expensive and impractical for use alone in an optimization setting. This section presents new approximations for fast statistical static timing analysis (FASSTA). This allows us to quickly evaluate costs of new gate assignments in subcircuits in the body of the optimization algorithm. The two operations needed in static timing analysis are sum and max. The FASSTA engine relies on the point values for means and variances of delays calculated in FULLSSTA rather than the complete discrete pdf representations.

We start with two normally distributed independent random variables *A* and *B* with expected values $\mu_A$ and $\mu_B$ and with variances $\sigma_A^2$ and $\sigma_B^2$ respectively. Let random variable *C* be the sum of *A* and *B*. The mean and variance of C are given by:

$$\mu_C = \mu_A + \mu_B \quad , \quad \sigma_C^2 = \sigma_A^2 + \sigma_B^2$$

To calculate the max, we shall expand on the formulation in [22]. We use the following notation:

$$\varphi(x) = \frac{1}{\sqrt{2\pi}} e^{\frac{-x^2}{2}}$$

$$\Phi(x) = \int_{-\infty}^{x} \varphi(t)\, dt$$

$$a^2 = \sigma_A^2 + \sigma_B^2$$

$$\alpha = \frac{(\mu_A - \mu_B)}{a}$$

The first two moments of *max(A,B)* are given by

$$\upsilon_1 = \mu_A \Phi(\alpha) + \mu_B \Phi(-\alpha) + a\varphi(\alpha) \quad (1)$$

$$\upsilon_2 = (\mu_A^2 + \sigma_A^2)\Phi(\alpha) + (\mu_B^2 + \sigma_B^2)\Phi(-\alpha) + (\mu_A + \mu_B) a\varphi(\alpha) \quad (2)$$

The variance of *max(A,B)* is given by

$$Var_{\max(A,B)} = \upsilon_2 - \upsilon_1^2 \quad (3)$$

These formulae cannot be evaluated directly because the integrals do not have analytical expressions and are

Proceedings of the Design, Automation and Test in Europe Conference and Exhibition (DATE'05)
1530-1591/05 $ 20.00 IEEE

expensive to compute. We show next how they can be avoided altogether. We reformulate the integral:

$$\Phi(x) = \int_{-\infty}^{x} \varphi(t)\,dt$$

$$\Phi(x) = \int_{-\infty}^{0} \varphi(t)\,dt + \int_{0}^{x} \varphi(t)\,dt$$

$$\Phi(x) = \frac{1}{2} + \frac{1}{2}\,erf\!\left(\frac{x}{\sqrt{2}}\right)$$

where *erf* denotes the error function. To calculate the error function, we use the following quadratic approximation [23] which is accurate to two decimal places

$$\frac{1}{2}\,erf\!\left(\frac{x}{\sqrt{2}}\right) \approx \begin{cases} 0.1x(4.4-x) & 0 \le x \le 2.2 \\ 0.49 & 2.2 < x < 2.6 \\ 0.50 & x \ge 2.6 \end{cases}$$

We also note that the error function is odd:

$$erf(-x) = -erf(x)$$

These formulae give us a quick method to approximate the error function for any value. We substitute this approximation in (1) and (2). We note that if

$$\alpha = \frac{(\mu_A - \mu_B)}{a} \ge 2.6 \qquad (5)$$

then

$$\Phi(\alpha) \approx 1,\ \Phi(-\alpha) \approx 0,\ \varphi(\alpha) \approx 0$$

and we have

$$\upsilon_1 \approx \mu_A\ ,\ \upsilon_2 \approx \mu_A^2 + \sigma_A^2$$

which gives

$$Mean_{\max(A,B)} \approx \mu_A\ ,\ Var_{\max(A,B)} \approx \sigma_A^2.$$

Similarly, for

$$\alpha = \frac{(\mu_A - \mu_B)}{a} \le -2.6 \qquad (6)$$

we get

$$Mean_{\max(A,B)} \approx \mu_B\ ,\ Var_{\max(A,B)} \approx \sigma_B^2$$

We observed that in the vast majority cases, one of (5) or (6) would apply obviating need for any calculation for max, while in other cases the approximations above provide quick estimates. These formulae assume independence of random variables which does not always hold. However, this approach emphasizes speed while retaining a reasonable degree of accuracy for small subcircuits. We stress that this approach is only used for the inner loop of the optimizations, while the outer loop relies on the more accurate discrete pdfs manipulation approach that can track correlations due to reconvergent paths using Principal Component Analysis [17] or other methods as long as runtime is managed appropriately.

### 4.4 Statistical critical path identification

As was pointed out in section 2.1, circuit optimization engines typically focus their effort on the critical or WNS path to improve the performance of the circuit. This section describes how we extend this concept to trace the Worst Negative Statistical Slack (WNSS) path in a circuit.

Consider a circuit consisting of 6 gates such as the one shown in Fig. 3. The first number in the parenthesis represents the statistical mean of delay for that arc while the second one represents the standard variation. We wish to determine the critical path with the biggest contribution to the variance at the output of node X. We note that, unlike the deterministic case, one cannot simply pick the input with the higher mean or variance to determine which input is most responsible for the variance at the output. This is due to the non-linearity of the statistical max operation where all inputs contribute to the output max.

We proceed to solve this problem by considering the sensitivity of the variance at the output of a node with respect to the inputs as follows. Starting from a given gate, we compare its inputs pair-wise. If either of (5) or (6) are satisfied, then we pick the input with the higher mean as clearly having the dominant influence on the output of this gate. If neither of these equations is satisfied, we compare

$$\frac{\partial Var_{\max(A,B)}}{\partial \mu_A} \quad \text{versus} \quad \frac{\partial Var_{\max(A,B)}}{\partial \mu_B}$$

Our justification for taking the partial derivatives with respect to the means of the delays is that the variances have a random component not under our direct control.

One approach to obtaining these sensitivities is to differentiate (3) directly. We found the resultant expressions to be complex and would require expensive floating-point computations. Instead, we chose to use an approximation for differentiation as follows. Rewriting

$$Var_{\max(A,B)} = f(\mu_A, \mu_B, \sigma_A, \sigma_B)$$

We use a forward finite-difference formula to approximate the partial derivative:

$$\frac{\partial Var_{\max(A,B)}}{\partial \mu_A} \approx$$

$$\frac{f(\mu_A + h, \mu_B, \sigma_A + g, \sigma_B) - f(\mu_A, \mu_B, \sigma_A, \sigma_B)}{h}$$

for small h.

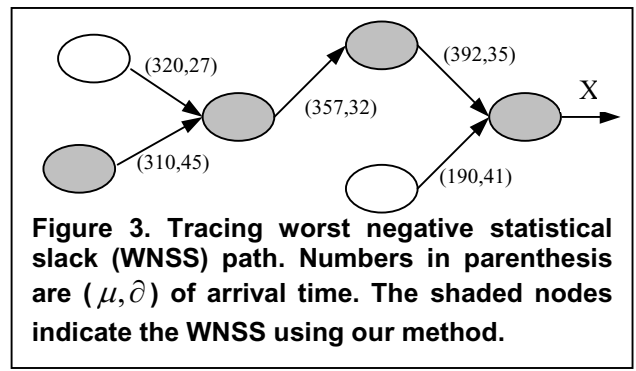

**Figure 3. Tracing worst negative statistical slack (WNSS) path. Numbers in parenthesis are ($\mu, \partial$) of arrival time. The shaded nodes indicate the WNSS using our method.**

We used values for h of the order of 1% of the mean. It should be noted that $\mu$ and $\sigma$ along a given path are correlated and one cannot expect to change one value without the other being impacted. The change in $\sigma_A$ that can result out of altering $\mu_A$ is indicated by g. We also note that it is impossible in general to determine g accurately as the relationship between $\mu$ and $\sigma$ along a given path is governed by a combination of gate performance variations inversely proportional to their dimensions as well unsystematic random variations that are unpredictable. For purposes of ranking inputs, the following linear approximation linking these two was found to be adequate:

$$g \approx \Delta\sigma \approx c\Delta\mu$$

We used values for c equal to those assumed to relate mean delay through a gate to its variance.

### 4.5 Subcircuit extraction and ranking

For every gate being evaluated for resizing, our algorithm extracts a subcircuit around this gate based on a user-controlled depth. We have found that using two levels of transitive fanins and fanouts is sufficiently accurate without being too costly to evaluate. For every available size for this gate, we use FASSTA to calculate mean and variance of delay at the outputs of this subcircuit. In order to rank the the relative merits of gate sizing in this subcircuit quickly, we use the following cost function. For all outputs of the subcircuit $O_1..O_n$, we calculate a weighted sum of mean and standard variation:

$$\text{Cost}(O_i) = \mu_i + \lambda\sigma_i \quad (7)$$

where $\lambda$ is a user-specified weight multiplier that ranks relative importance of minimizing standard variation against mean of delay. By choosing higher values for $\lambda$, the user can place more emphasis on variance reduction. We provide more analysis on effect of varying $\lambda$ in the conclusions section at the end of the paper. The cost of the subcircuit is given by the maximum of $\text{Cost}(O_i)$ across all outputs. We then pick the gate size that minimizes subcircuit cost across all gate sizes for candidate gate.

### 5. Experimental results

The proposed approach was implemented in Java and run on an Intel PC running at 2.53 GHz. We tested the algorithm on various circuits from the ISCAS benchmarks and various sized ALU circuits. The circuits were first synthesized using Design Compiler [24] using an industrial 90nm lookup-table based standard cell library with 6-8 sizes per gate type. In line with other researchers, we added variations to the gate delays based on [25,26]. Two variations components were added to the gate delays: one proportional to delay through gate and another random source corresponding to unsystematic manufacturing variations.

Table 1 shows the results of our optimization. The ratio of $\sigma$ to $\mu$ obtained by optimizing for mean delay is shown in the first column entitled original. We then ran our algorithm at various values for $\lambda$ (7). Results are shown for optimization under two different values for $\lambda$, 3 and 9. We observed that increasing $\lambda$ any further could not yield further reduction in variance in general though the highest value for $\lambda$ was different for different circuits. This is due to the unsystematic variations whose effects cannot be eliminated. Fig. 4 below shows a plot of $\mu$ against $\sigma$ for various values of $\lambda$ for circuit C432.

Several observations can be made from these results. Our algorithm consistently reduces the standard variation while increasing mean delay and area. This behavior is expected since our algorithm favors bigger gate sizes that reduce the variance of delay across them. The algorithm's focus on minimizing variance also causes it to upsize gates near the outputs to reduce the overall variance at circuit's output. This is done even if that path does not have the highest mean delay which is in contrast to a worst mean-delay optimizer which would not upsize such gates. This increases overall delay due to higher loading slowing predecessor gates.

Another important observation is that the number of gates along a timing path is inversely proportional to the variance along that path and the ability to optimize it away. Paths with a shorter number of gates tend to be more susceptible to variations. The smaller ALU circuits exhibit significant variations as a percentage of their mean. Our algorithm can reduce this variation substantially but at a higher increase in area. On the other hand, circuit C6288 which is a 16x16 bit multiplier has the longest depth of any of the circuits in the table. We note that it has the lowest improvement due to its already low $\sigma$ to $\mu$ ratio.

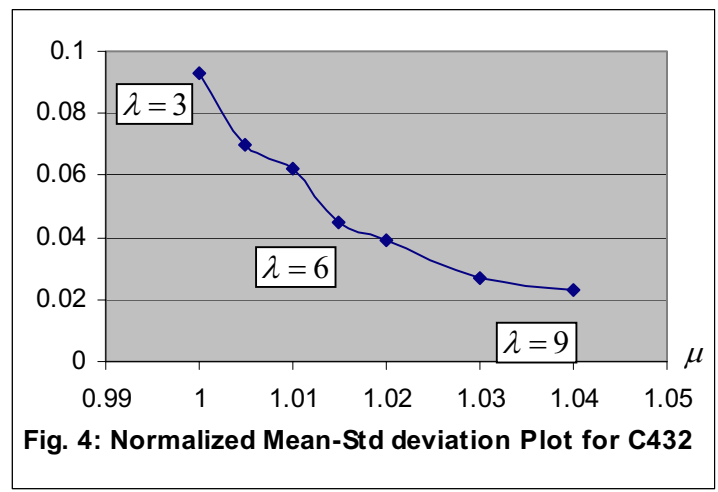

Fig. 4: Normalized Mean-Std deviation Plot for C432

**Table 1. Results of our approach on ISCAS Benchmarks**

| Circuit Name | Gates | Original $\sigma/\mu$ | $\lambda = 3$ | | | | | $\lambda = 9$ | | | | |
|---|---|---|---|---|---|---|---|---|---|---|---|---|
| | | | $\Delta\mu$ | $\Delta\sigma$ | $\sigma/\mu$ | $\Delta A$ | Run Time (Minutes) | $\Delta\mu$ | $\Delta\sigma$ | $\sigma/\mu$ | $\Delta A$ | Run Time (Minutes) |
| alu1 | 234 | 0.124 | +4 % | -54 % | 0.055 | +16 % | 1.5 | +6 % | -80 % | 0.023 | +24 % | 1.6 |
| alu2 | 161 | 0.147 | +3 % | -71 % | 0.041 | +14 % | 1.3 | +4 % | -86 % | 0.020 | +29 % | 1.4 |
| alu3 | 215 | 0.127 | +7 % | -61 % | 0.046 | +16 % | 1.5 | +9 % | -75 % | 0.029 | +25 % | 1.7 |
| c432 | 203 | 0.093 | +2 % | -58 % | 0.038 | +11 % | 1.6 | +4 % | -75 % | 0.022 | +21 % | 1.7 |
| c499 | 381 | 0.077 | +5 % | -63 % | 0.027 | +13 % | 1.5 | +8 % | - 76 % | 0.017 | +21 % | 1.8 |
| c880 | 301 | 0.092 | +4 % | -57 % | 0.038 | +17 % | 1.5 | +5 % | -79 % | 0.018 | +23 % | 1.7 |
| c1355 | 378 | 0.081 | +5 % | -63 % | 0.057 | +13 % | 1.7 | +7 % | -71 % | 0.022 | +19 % | 1.9 |
| c1908 | 563 | 0.076 | +3 % | -44 % | 0.041 | +7 % | 3.7 | +4 % | -71 % | 0.021 | +16 % | 3.8 |
| c2670 | 820 | 0.068 | +2 % | -42 % | 0.039 | +11 % | 9.8 | +7 % | -76 % | 0.015 | +18 % | 9.1 |
| c3540 | 1245 | 0.062 | +4 % | -56 % | 0.026 | +12 % | 14.7 | +8 % | -70 % | 0.017 | +21 % | 13.1 |
| c5315 | 2318 | 0.043 | +2 % | -36 % | 0.027 | +12 % | 36 | +7 % | - 68 % | 0.013 | +15 % | 34 |
| c6288 | 2980 | 0.021 | +1 % | -28 % | 0.015 | +5 % | 44 | +2 % | - 47 % | 0.011 | +9 % | 41 |
| c7552 | 2763 | 0.043 | +2 % | -50 % | 0.021 | +11 % | 31 | +4 % | - 66 % | 0.014 | +17 % | 33 |

## 6. Concluding Remarks

We introduced a new concept of a worst negative statistical slack path and derived a procedure for tracing and optimizing such paths. In the process, we also derived a new approximation for the max operation on random variables for use in circuit optimization. Our approach allows us to steer the optimization process towards different mean-variance goals. The significance of this work is that it can be used during design cycle to increase tolerance for the effects of manufacturing variations by trading off circuit delay and area requirements for reduced timing variance with user controlled weights. We demonstrated fidelity of our approach on ISCAS benchmarks with consistent variance reduction in exchange for moderate increases in area and low increases in mean delays.